\documentclass{aipproc}

\layoutstyle{6x9}  

\usepackage{graphicx,epsf}   
\usepackage{psfrag}

\def\al{\!\!&\!\!}
\def\be{\begin{equation}}
\def\ee{\end{equation}}
\def\bea{\begin{eqnarray}}
\def\eea{\end{eqnarray}}
\newcommand{\bdm}{\begin{displaymath}}
\newcommand{\edm}{\end{displaymath}}

\newcommand{\co}{\; ,}

\begin{document}
\title{On the dispersion theory of $\pi\pi$ scattering}
      
\author{H. Leutwyler}{email={leutwyler@itp.unibe.ch},
  address={Institute for Theoretical Physics, University of Bern,
  Sidlerstr.~5, CH-3012 Switzerland}}

\begin{abstract}Recent developments in low energy pion physics are reviewed,
  emphasizing the strength of dispersion theory in this context. As an
  illustration of the method, I discuss some consequences of the forward
  dispersion relation obeyed by the isoscalar component of the scattering
  amplitude.

  \vspace{0.3cm} \begin{center} Contribution to the proceedings of 
     the workshop {\it Quark Confinement and the Hadron Spectrum VII}, 

Ponta Delgada, Azores Islands, Portugal, September 2006\end{center}

\end{abstract}

\maketitle

\vspace{1em}Pions play a crucial role whenever the strong interaction is
involved at low energies -- the Standard Model prediction for the muon
magnetic moment provides a good illustration. My talk dealt with the
remarkable theo\-re\-ti\-cal progress made in low energy pion physics in
recent years. 

In the first part, I discussed the current theoretical and experimental
knowledge of the $S$-wave $\pi\pi$ scattering lengths ($a_0^0$, $a_0^2$) in
some detail, because these play a central role.  Simulations of QCD on a
lattice now reach sufficiently small quark masses for a meaningful
extrapolation to the values of physical interest to become possible
\cite{MILC,Del Debbio et al}. Chiral perturbation theory not only describes
the dependence on the quark masses, but also allows one to calculate the
leading finite volume effects \cite{Colangelo Duerr Haefeli}. The values
obtained for the coupling constants $\ell_3$ and $\ell_4$ are consistent with
the estimates given in \cite{GL 1984,CGL}. Since these control the leading
corrections to the low energy theorems for $a_0^0$ and $a_0^2$, the lattice
results at the same time provide a rough check of the remarkably precise
predictions for these quantities obtained in \cite{CGL} (in the following,
this paper is referred to as CGL). The exotic scattering length $a_0^2$ can
also be extracted directly from the volume dependence of the energy levels
occurring on the lattice. The result obtained in \cite{Beane} is in good
agreement with the prediction as well. The current state of our knowledge
concerning $a_0^0$ and $a_0^2$ is briefly summarized in \cite{Krakow}.

The second part of the talk covered recent results established with dispersive
methods. The upshot of this development is that, in the threshold region, the
$\pi\pi$ scattering amplitude is now known to an amazing degree of accuracy
\cite{CGL}. In particular, we know how to calculate mass and width of the
lowest resonance of QCD \cite{CCL}. The actual uncertainty in the pole
position is smaller than the estimate given in the 2006 edition of the Review
of Particle Physics \cite{PDG 2006}, by more than an order of magnitude. The
progress made in this field heavily relies on the fact that the dispersion
theory of $\pi\pi$ scattering is particularly simple: the $s$-, $t$- and
$u$-channels represent the same physical process.  As a consequence, the
scattering amplitude can be represented as a dispersion integral over the
imaginary part and the integral exclusively extends over the physical region
\cite{Roy}. The projection of the amplitude on the partial waves leads to a
dispersive representation for these, the Roy equations. For a detailed
discussion, I refer to \cite{ACGL}.

In the present article, I wish to explain the essence of the dispersive
approach, avoiding technical machinery as much as possible. Although the Roy
equations represent an optimal and comprehensive framework for the low energy
analysis of the $\pi\pi$ scattering amplitude, the main points can be seen in
a simpler context: forward dispersion relations \cite{KPY}. More specifically,
I consider the component of the scattering amplitude with $s$-channel isospin
$I=0$, which I denote by $T^0(s,t)$. It satisfies a twice subtracted fixed-$t$
dispersion relation in the variable $s$. In the forward direction, $t=0$, this
relation reads \bea
\label{eqFDR}
\mbox{Re}\,T^0(s,0)\al=\al c_0 + c_1\, s+ \frac{s(s-4M_\pi^2)}{\pi}\,
P\hspace{-0.3em}\int_{4M_\pi^2}^\infty\frac{dx
  \;\mbox{Im}\,T^0(x,0)} {x\,(x-4M_\pi^2)\,(x-s)}+\\
\al+\al\frac{s(s-4M_\pi^2)}{\pi} \int_{4M_\pi^2}^\infty\frac{dx
  \;\{\mbox{Im}\,T^0(x,0)-3\,\mbox{Im}\,T^1(x,0)+ 5\,\mbox{Im}\,T^2(x,0)\}}
{3\,x\,(x-4M_\pi^2)\,(x+s-4M_\pi^2)}\,.\nonumber\eea The symbol $P$ indicates
that the principal value must be taken. The first integral accounts for the
discontinuity across the right hand cut, while the second represents the
analogous contribution from the left hand cut, where the components of the
scattering amplitude with $I=1,2$ also show up. According to the optical
theorem, the imaginary part of the forward scattering amplitude represents the
total cross section: in the norma\-li\-zation of \cite{ACGL}, we have
$\mbox{Im}\,T^I(s,0)=\sqrt{s(s-4M_\pi^2)}\;\sigma^I_{tot}(s)$. The
subtraction constants are also determined by physical quantities -- the
$S$-wave scattering lengths:  \be
c_0+c_1\, s = 32\,\pi
\left\{a_0^0+(2a_0^0-5a_0^2)\,\frac{s-4M_\pi^2}{12M_\pi^2}\right\}\,.\ee

A dispersion relation of the above type also holds for other processes. What
is special about $\pi\pi$ is that the contribution from the crossed channels
can be expressed in terms of observable quantities -- total cross sections in
the case of forward scattering.

The right hand side of equation (1) can be evaluated with the available
representations of the scattering amplitude. The contribution from the left
hand cut is dominated by the $\rho$-meson, which generates a pronounced peak
in the total cross section with $I=1$. This contribution is known very
accurately from $e^+e^-\rightarrow\pi^+\pi^-$ and
$\tau^\pm\rightarrow\nu\hspace{0.05cm} \pi^\pm\pi^0$. Since the channel with
$I=2$ is exotic, it does not contain any resonances and -- at low energies --
only generates a minor correction. In the physical region, $s> 4M_\pi^2$, the
entire contribution from the crossed channels is a smooth function that varies
only slowly with the energy. Note, however, that this contribution is by no
means small.
\begin{figure}[thb]
\includegraphics[width=4.3cm,angle=-90]{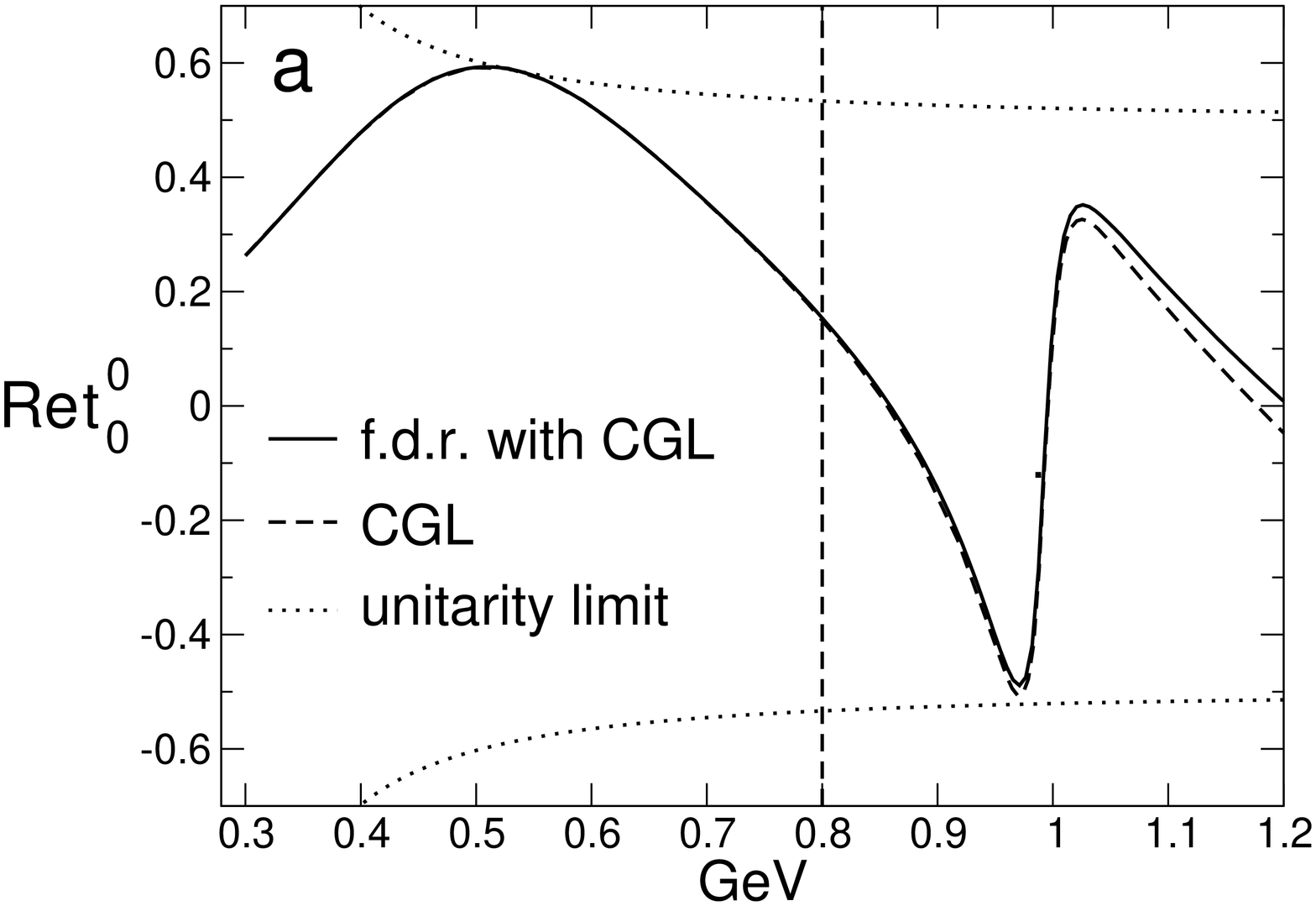}\hspace{-0.6cm}
\includegraphics[width=4.3cm,angle=-90]{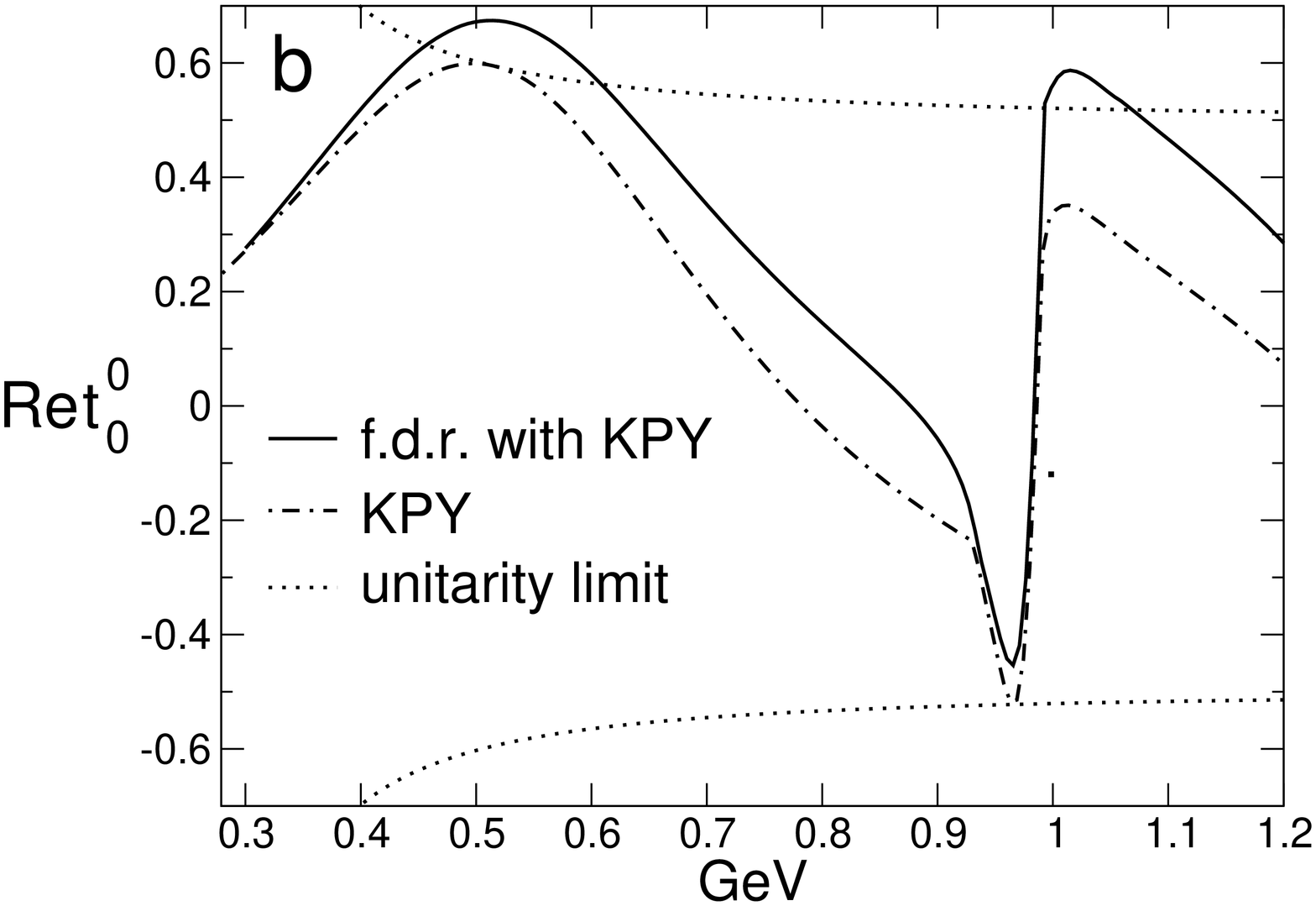}\hspace{-0.6cm}
\includegraphics[width=4.3cm,angle=-90]{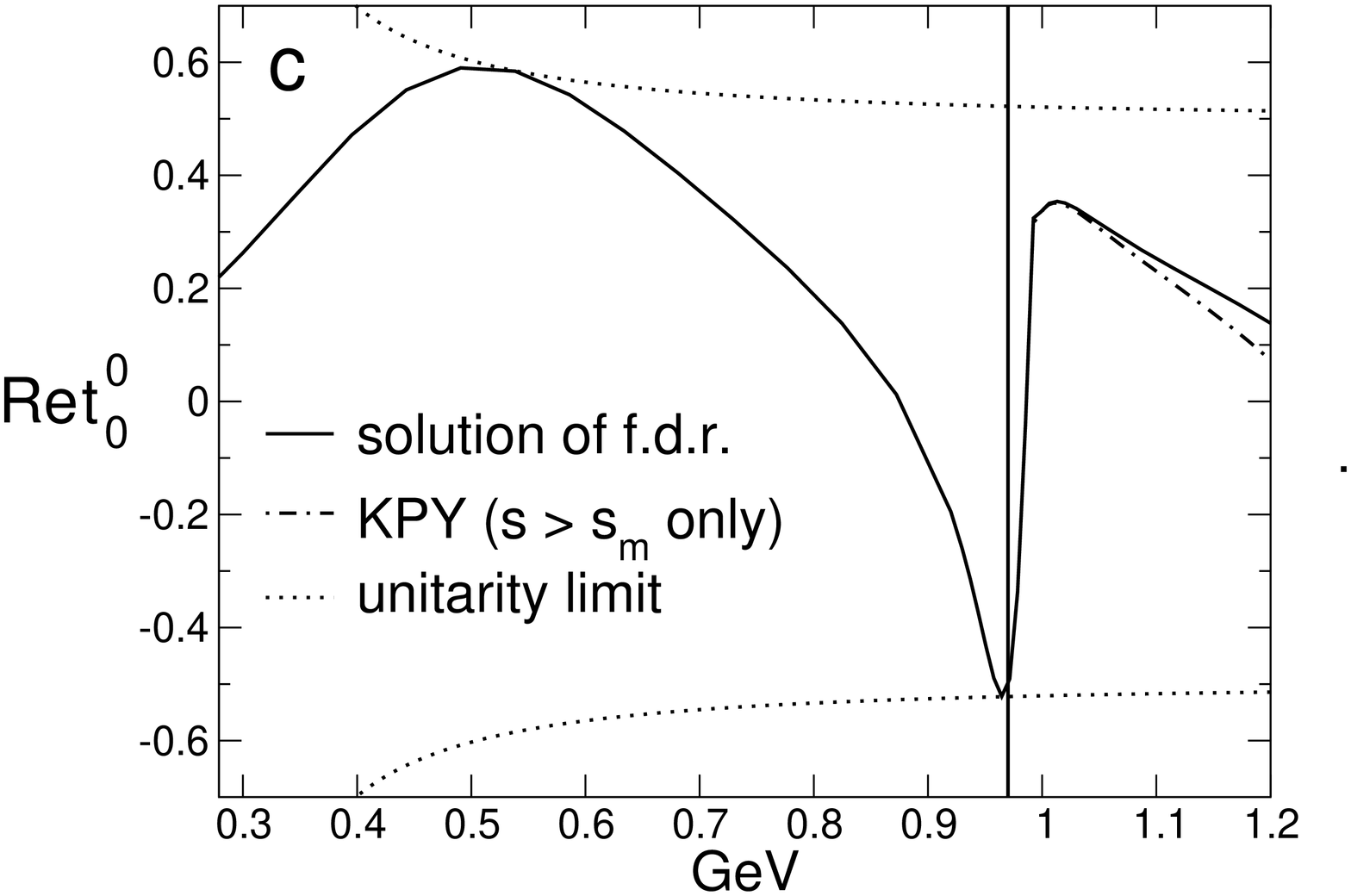}
\caption{\label{figret00}Real part of the isoscalar $S$-wave from forward
  dispersion relation. }
\end{figure}

The angular momentum barrier suppresses the higher partial waves: at low
energies, the first term in the partial wave decomposition \be\label{eqPWE}
\mbox{Re}\,T^0(s,0)/(32\pi)=\mbox{Re}\,t^0_0(s)+5\,\mbox{Re}\,t^0_2(s)+
\ldots\co\ee represents the most important contribution. In the vicinity of
the threshold, where the contribution from the $D$-wave is small, the
dispersion relation (\ref{eqFDR}) thus essentially determines the real part of
the isoscalar $S$-wave. For brevity, I refer to this wave as $S^0$.

\begin{figure}[thb]
\vspace{-1em}\includegraphics[width=10cm,angle=-90]{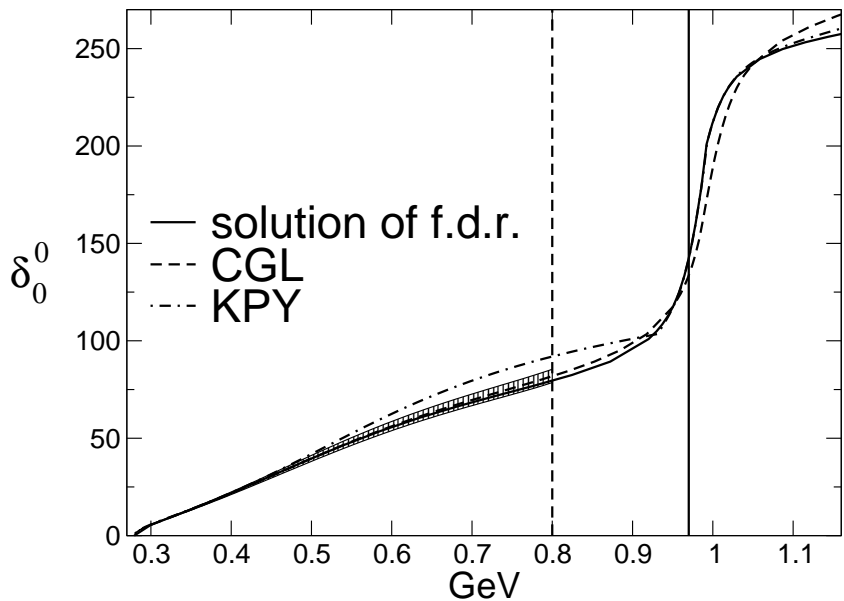}
\caption{\label{figdelta00}Phase of the isoscalar
  $S$-wave.}    
\end{figure}

\begin{figure}[thb]
\includegraphics[width=6.2cm,angle=-90]{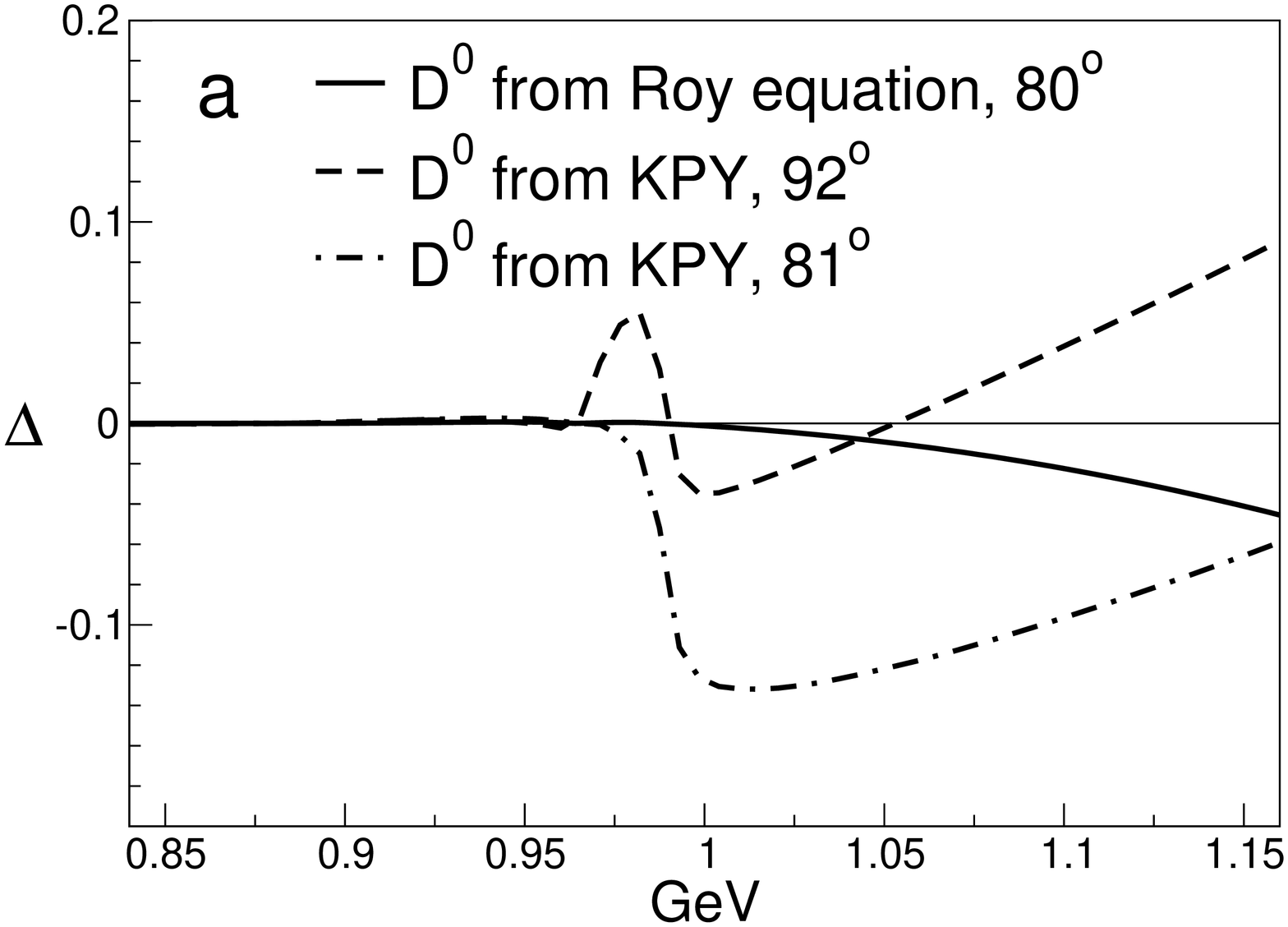}\hspace{-0.5cm}
\includegraphics[width=6.2cm,angle=-90]{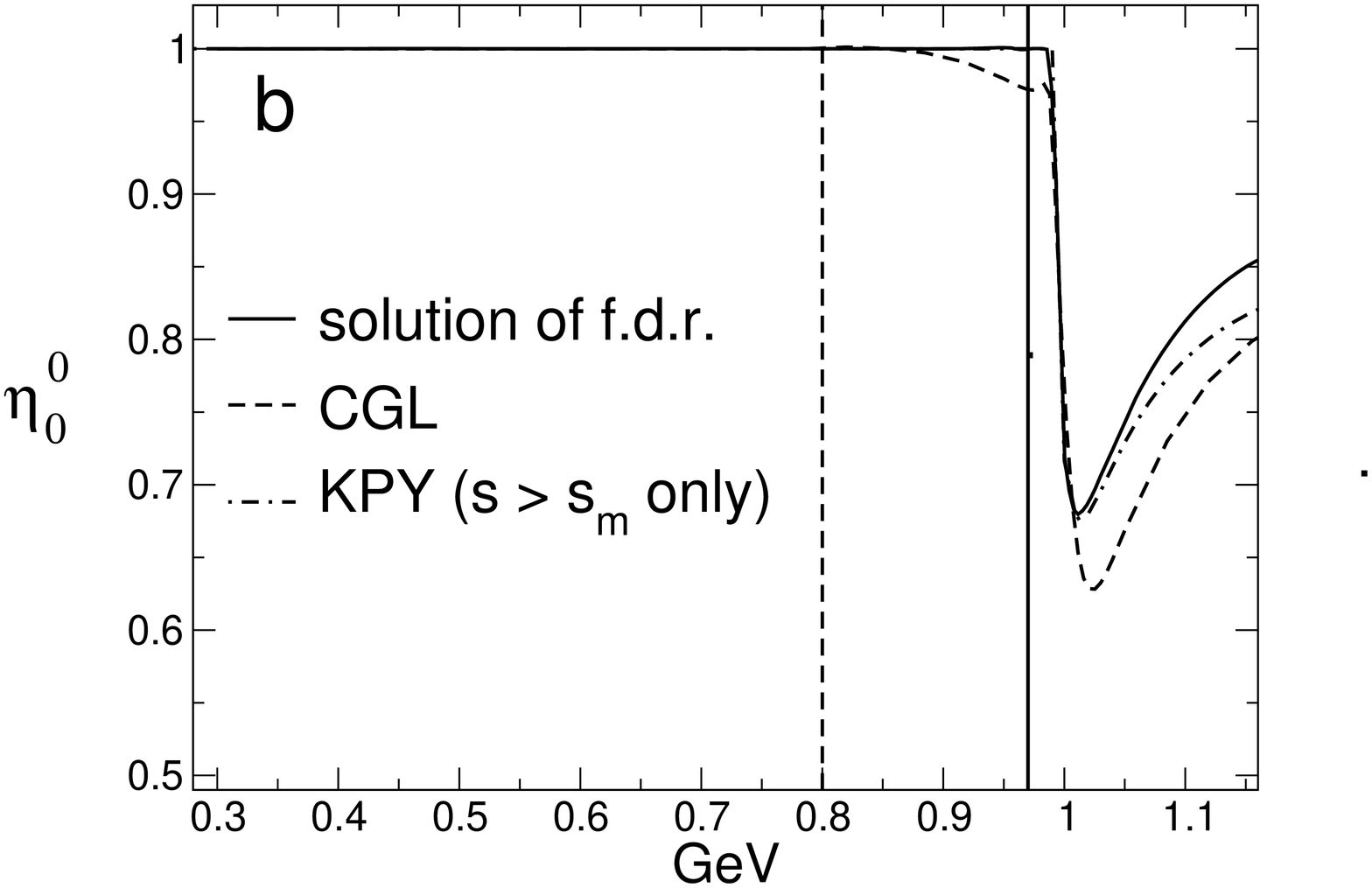}
\caption{\label{figeta00}Violation of f.d.r.\ for $T^0$ and elasticity of the
  isoscalar 
  $S$-wave.}    
\end{figure}

Figure \ref{figret00}a is based on the representation of the scattering
amplitude in CGL, where the low energy behaviour of the $S$- and $P$-wave
phase shifts was determined by solving the Roy equations below 800 MeV. That
calculation required input for (a) the imaginary parts of the higher partial
waves, (b) the imaginary parts of the $S$- and $P$-waves above 800 MeV and (c)
the $S$-wave scattering lengths. For (a) and (b), we relied on the literature
\cite{literature}, while for (c), we used the low energy theorems of chiral
perturbation theory. The Roy equations then yield an approximate
representation for the real parts of all partial waves, throughout their
region of validity. Together with the input used for the imaginary parts, this
also fixes the phase shifts and elasticities in that region. The dashed lines
in figures \ref{figret00}a, \ref{figdelta00} and \ref{figeta00}b are
calculated in this way. I emphasize that, above 800 MeV, these curves amount
to an extrapolation.  The uncertainties in the representation of the
scattering amplitude are discussed in detail in CGL, but are not shown in the
figures, which are calculated with the central values. In the threshold
region, the uncertainties are tiny, but they grow with the energy. In
particular, Figure \ref{figeta00}b shows that the extrapolation overestimates
the inelasticity in the region between 800 MeV and $2M_K$ -- in reality, a
significant amount of inelasticity only arises when the $K\bar{K}$ channel
opens.

In order to demonstrate that the representation of the scattering amplitude in
CGL is consistent with equation (\ref{eqFDR}), I evaluate the function
Re$\,T^0(s,0)$ with it and remove the $D^0$-wave, setting
Re$\,t^0_0(s)|_{f.d.r.}= \mbox{Re}\, T^0(s,0)/(32\pi)-5\,\mbox{Re}\,t^0_2(s)$
and thereby ignoring partial waves with $\ell \geq 4$. Figure \ref{figret00}a
shows that, below 800 MeV, the f.d.r.\ is indeed obeyed very well. No wonder:
if the partial waves satisfy the Roy equations, then the sum over
\underline{all} of these automatically obeys the f.d.r. The difference between
the two curves arises from the neglected higher partial waves, which become
more important if the energy is increased.

Figure \ref{figret00}b shows the result obtained if the representation in CGL
is replaced by the one in \cite{KPY} (in the following, this paper is referred
to as KPY). Visibly, the forward dispersion relation is not obeyed well, but
this is to be expected: that parametrization does not rely on dispersion
theory, but represents a phenomenological fit to data sets that are subject to
large errors. Note also that the figure does not show the uncertainties of the
partial wave analysis in KPY, which are considerable: the difference between
the two curves is covered by these.

As discussed in \cite{KPY,KPY Madrid}, the forward dispersion relations can be
used to improve the partial wave representation. In the following, I apply the
method of \cite{ACGL}, which is very suitable for the purpose: it suffices to
replace the Roy equation for Re$\,t^0_0$ by the f.d.r.\ for Re$\,T^0$. If the
subtraction constants and all partial waves except $t^0_0$ are treated as
known, the problem may be given the following mathematical form:

{\it Choose a ``matching point'' $s_m$ and prescribe the function
Im$\,t^0_0(s)$ above $s_m$ as well as the elasticity $\eta^0_0(s)$ below
$s_m$. Find solutions of equation (\ref{eqFDR}) for $s<s_m$ that
respect the unitarity relation between Re$\,t^0_0$, Im$\,t^0_0$ and
$\eta^0_0$.} 

In the framework of the Roy equations, this problem was discussed in detail in
\cite{ACGL}. If the matching point is taken below the energy where the phase
goes through $90^\circ$, then there is exactly one solution. This is the
situation considered in CGL, where the matching point and the central value of
the phase at that point were set equal to 800 MeV and $82.3^\circ$,
respectively.  The dashed lines in figures \ref{figret00}a, \ref{figdelta00}
and \ref{figeta00}b represent the resulting unique solution. 

Next, I observe that, near 970 MeV, the representations given for $S^0$ in KPY
and CGL are very similar -- in either case, the real part reaches the lower
unitarity limit in the immediate vicinity of that energy. For this reason, I
now discuss the situation for the case where the matching point is taken at
$\sqrt{s_m}=970$ MeV (full vertical line) and where the input for the
imaginary parts is taken from KPY: Above 1.42 GeV, I use the Regge
parametrization of the forward scattering amplitudes given in that reference.
At lower energies, the partial wave decomposition is used. With the exception
of $S^0$ below 970 MeV, all of the partial waves are taken from KPY (central
values of the parameters, throughout).  Finally, below 970 MeV, the elasticity
of $S^0$ is set equal to 1, while the phase is left open -- the f.d.r.\ is
used to determine it. In order to be able to compare the result with the one
obtained with CGL, I keep the subtraction constants fixed at the central
values in CGL.  The range used for $a_0^0,a_0^2$ in KPY  is consistent with
that.

With this input, the value of the phase at the matching point is $142^\circ$.
As discussed in detail in \cite{ACGL}, the mathematical problem posed above
leads to a curiosity if the phase at the matching point exceeds $90^\circ$:
the solution of the forward dispersion relation then fails to be unique --
there is an entire family of solutions. In the present case, there is a
one-parameter family, which may be labeled with the value of the phase
somewhere below the matching point, at 800 MeV, for example. In particular, we
may select the solution for which the phase at 800 MeV agrees with the central
parametrization in KPY, $\delta_0^0(800\,\mbox{MeV})= 92^\circ$.

This particular solution is physically unacceptable, however, because it
contains a strong cusp at the matching point. An enlargement of the region
around this point is shown in figure \ref{figeta00}a, where the difference
$\Delta$ between the left and right hand sides of equation (\ref{eqFDR}) is
plotted as a function of the energy.  By construction, the difference vanishes
below the matching point -- within the accuracy to which the solutions are
worked out.  The cusp manifests itself as a spike in the vicinity of the
matching point.  The occurrence of a cusp is a generic feature of the manifold
of solutions to the mathematical problem specified above. The amplitude of
the cusp decreases if the phase at 800 MeV is lowered. There is a unique
solution for which a cusp does not occur, in the sense that the first
derivative is continuous at the matching point. This solution is reached if
the phase at 800 MeV is lowered to about $81^\circ$. Figure \ref{figeta00}a
shows that the violation of the f.d.r.\ is then postponed to $K\bar{K}$
threshold, but it persists: with the input specified above, equation
(\ref{eqFDR}) does not have a physically acceptable solution.

The problem originates in $D^0$, the isoscalar $D$-wave -- indeed, in KPY, the
parametrization used for the elasticity of this wave is mentioned as a
potential culprit. Around $K\bar{K}$ threshold, the real part of $D^0$ is by
no means negligible and it is essential that the representation used for it is
consistent with analyticity. In this regard, there is a difference between the
mathematical problem specified above and the Roy equations, where the input of
the calculation exclusively involves imaginary parts. As was noted already in
the pioneering work of Basdevant, Froggatt and Petersen \cite{Basdevant
  Froggatt Petersen}, dispersion theory imposes very strong constraints on the
low energy behaviour of the higher partial waves.  The representation for
$D^0$ in KPY is not consistent with these constraints. In particular, phase
space strongly suppresses the inelasticity generated by the $K\bar{K}$ states:
$1-\eta^0_2$ only grows with the fifth power of the kaon momentum.

The problem encountered with the behaviour of the solutions around $K\bar{K}$
threshold disappears if the representation for $D^0$ is taken from the
solution of the Roy equation for Re$\,t^0_2$, retaining the parametrization of
KPY only for the other waves. Alternatively, the parametrization of $D^0$
given in \cite{Hyams} can be used -- this is less accurate, but barely makes
any difference around $K\bar{K}$ threshold.  The full lines in figures
\ref{figeta00}a and \ref{figeta00}b show that equation (\ref{eqFDR}) does now
admit physically acceptable solutions. As before, there is a one-parameter
family of solutions, which differ in the strength of the cusp at the matching
point. The particular solution shown is obtained by minimizing the difference
between the right and left hand sides of equation (\ref{eqFDR}) on the entire
interval from $2M_\pi$ to $2M_K$. For this solution, the cusp is too small to
be visible in the figures.

As was to be expected, the solution represents a compromise between the two
curves in figure \ref{figret00}b. Below $2M_K$, the difference between the
left and right hand sides of equation (\ref{eqFDR}) is less than $10^{-3}$,
but above 1050 MeV, the real part of the solution departs from the real part
of the parametrization in KPY, which is indicated as a dash-dotted line (shown
only above 970 MeV, where the imaginary part of that parametrization is used
as an input). The real part and the phase of the solution are displayed as
full lines in figures \ref{figret00}c and \ref{figdelta00}. At low energies,
the solution of the forward dispersion relation runs within the shaded region,
which represents the uncertainty band in CGL. The value of the phase at the
upper end is $\delta_0^0(800\,\mbox{MeV})=80^\circ$. This confirms one of the
main results in CGL: below 800 MeV, the solution is not sensitive to the
behaviour at high energies. On the other hand, above the matching point, the
phase very closely follows the parametrization of the $S^0$-wave used in the
input. As discussed in detail in \cite{ACGL}, in the context of the Roy
equations, this happens whenever the input stems from an analytic
parametrization of the partial wave in question.  As can be seen in figure
\ref{figeta00}b, shifting the matching point up has the effect of removing the
unitarity violation in the elastic region -- the output for $\eta_0^0$ now
differs appreciably from KPY only above 1050 MeV.

I conclude that -- in the specific framework used here, where all partial
waves except the isoscalar $S$-wave are treated as known -- the constraint
imposed on the scattering amplitude by the forward dispersion relation for
$T^0(s,t)$ is essentially equivalent to the Roy equation for Re$\,t^0_0(s)$.
Likewise, the Roy equations for Re$\,t^1_1(s)$ and Re$\,t^2_0(s)$ can be
replaced by the forward dispersion relations for $T^1(s,t)$ and $T^2(s,t)$. In
all three cases, the subtraction constants are determined by the $S$-wave
scattering lengths. The forward dispersion relations do not provide a handle
on the higher partial waves, but the Roy equations do. As we have pushed the
matching point up, the dispersive framework now correlates the value of the
phase at 800 MeV with the behaviour of the scattering amplitude above
$K\bar{K}$ threshold. In particular, if the representation in KPY represents a
good approximation above $2M_K$ and if the theoretical predictions for $a_0^0$
and $a_0^2$ are correct, then the phase at 800 MeV must be in the vicinity of
$80^\circ$.

In the above, I did not discuss the experimental information at all.  Instead,
I merely showed that -- in view of the sharp theoretical predictions for the
subtraction constants -- the properties of the amplitude above $K\bar{K}$
threshold very strongly constrain the behaviour at lower energies.  A more
systematic investigation is under way, which aims at extending the work
described in CGL to somewhat higher energies. The matching point can be pushed
up all the way to the limit of validity of the Roy equations. The price to pay
is that the contributions from the high energy region then become more
important. We are using a Regge representation for the asymptotic domain and
invoke experimental information as well as sum rules to pin down the residue
functions. A short account of this work is given in \cite{Beijing} and more
should soon be ready for publication, in particular also the application to
the electromagnetic form factor of the pion, for which an accurate
representation is needed in connection with the Standard Model prediction for
the magnetic moment of the muon.

It is a pleasure to thank the organizers of the meeting for their kind
hospitality during a very pleasant stay on the Azores islands. Also, I
acknowledge Irinel Caprini and Gilberto Colangelo for a most enjoyable and
fruitful collaboration and thank Hanqing Zheng for many informative
discussions. Also, I am indebted to Jos\'{e} Pel\'{a}ez and Francisco
Yndur\'{a}in for comments on the manuscript. 

\end{document}